\documentclass[prl,aps,amsfonts,twocolumn,floatfix,superscriptaddress,nofootinbib]{revtex4-1}

\usepackage{lineno,hyperref}
\modulolinenumbers[5]
\usepackage{graphicx,amssymb,color,amsmath}
\usepackage{natbib}
\usepackage{braket}
\usepackage{cases}
\usepackage{color}
\usepackage{mathtools}
\usepackage{caption}
\usepackage{subcaption}
\usepackage{multirow}
\usepackage[normalem]{ulem}

\definecolor{orange}{rgb}{1,0.5,0}

\begin{document}

\title{Time-reversed particle-vibration loops and nuclear Gamow-Teller response}

\author{Caroline Robin}
\email{carolr8@uw.edu}
\affiliation{Institute for Nuclear Theory, University of Washington, Seattle, WA 98195, USA.}
\affiliation{JINA-CEE, Michigan State University, East Lansing, MI 48824, USA.}

\author{Elena Litvinova}
\affiliation{Department of Physics, Western Michigan University, Kalamazoo, MI 49008-5252, USA}
\affiliation{National Superconducting Cyclotron Laboratory, Michigan State University, East Lansing, MI 48824, USA}

\date{\today}

\begin{abstract}
The nuclear response theory for charge-exchange modes in the relativistic particle-vibration coupling approach is extended to include for the first time particle-vibration coupling effects in the ground state of the parent nucleus.
In a framework based on the effective meson-nucleon Lagrangian, we investigate the role of such complex ground-state correlations in the description of Gamow-Teller transitions in $^{90}$Zr in both (p,n) and (n,p) channels.
The particle-vibration coupling effects are calculated without introducing new parameters.
We find that this new correlation mechanism is fully responsible for the appearance of the strength in the (n,p) branch. Comparison of our results to the available experimental data shows a very good agreement 
up to excitation energies beyond the giant resonance region when taking into account a phenomenological admixture of the isovector spin monopole transitions.
The parent-daughter binding-energy differences are also greatly improved by the inclusion of the new correlations.
\end{abstract}
\maketitle

\begin{figure}[h]
\centering{\includegraphics[width=\columnwidth] {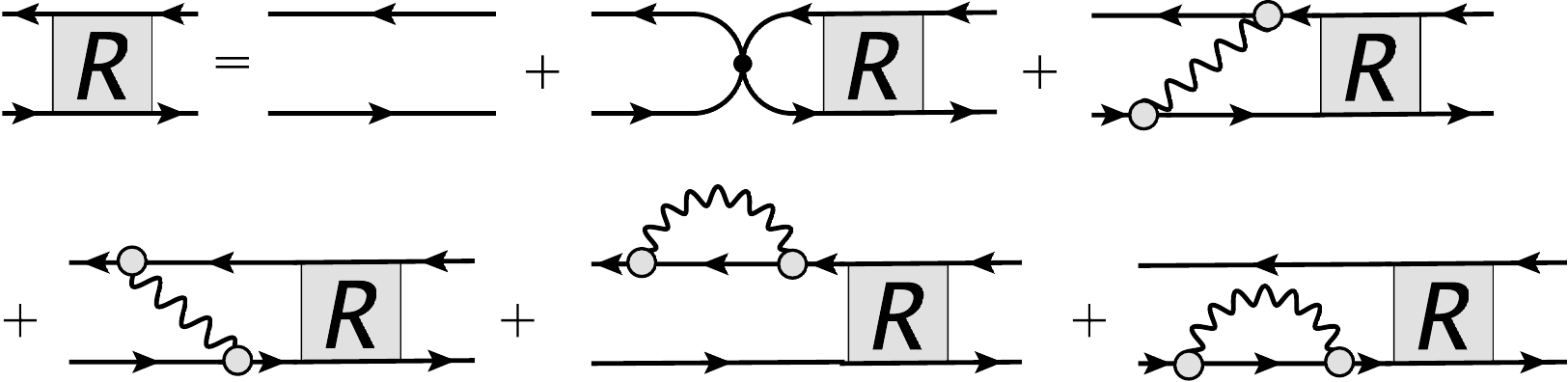}}  \hfill %
\caption{Graphical BSE for the $R = R^{(--)}$ component in the resonant TBA. Left- (right-)oriented solid lines denote one-nucleon propagators in particle (hole) states. 
The first two terms on the r.h.s correspond to pn-RRPA and the black circle represents the static isovector meson-exchange interaction (with both forward and backward contributions generating GSC$_{RPA}$). In the next terms weavy lines denote phonon propagators and gray circles the PVC amplitudes. These terms describe phonon exchanges between protons and neutrons as well as virtual emissions and reabsorption of phonons by nucleons.}
 \label{f:Phi_NOGSC} 
\end{figure}

\begin{figure}[h]
\centering{\includegraphics[width=\columnwidth] {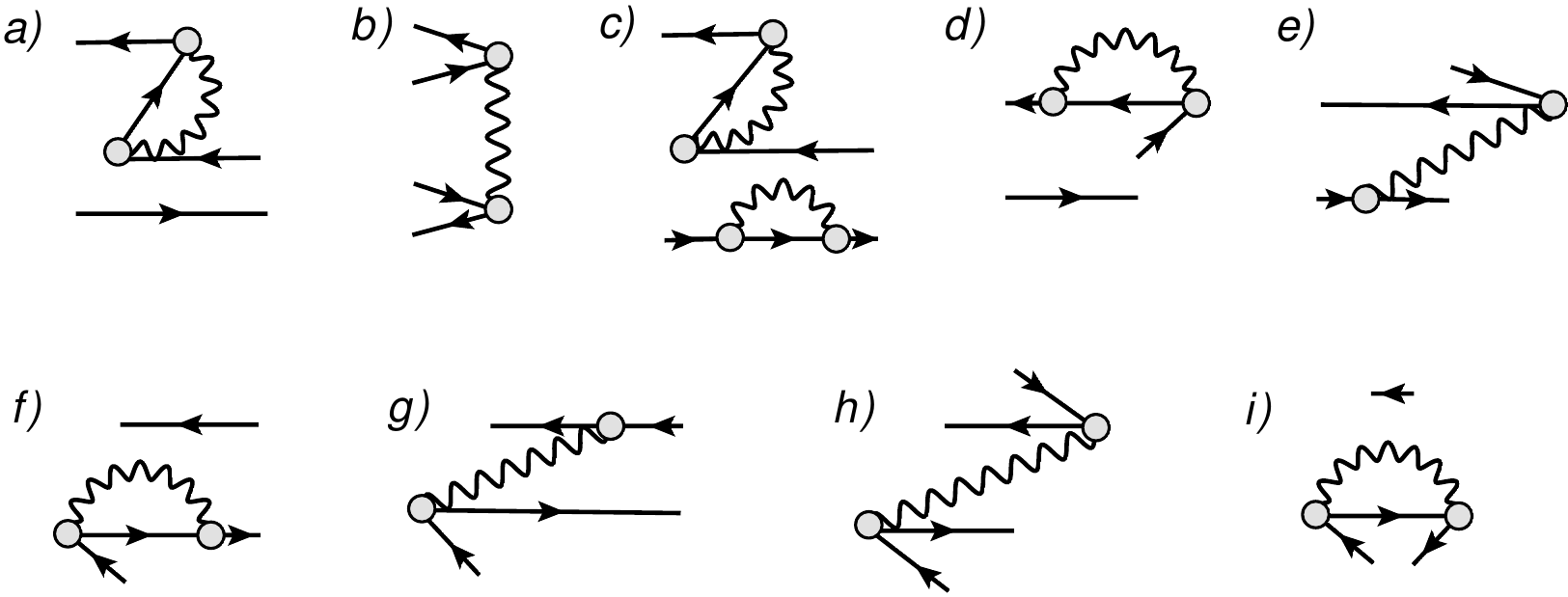}}  \hfill %
\caption{Some diagrams induced by GSC$_{PVC}$. Diagrams a),b),c) appear in the channel $R^{(--)}$ of the response, d),e) in $R^{(-+)}$, f),g) in $R^{(+-)}$, and h),i) in $R^{(++)}$.}
 \label{f:GSC} 
\end{figure}

{\it Introduction.} 
Charge-exchange transitions in nuclei, in particular Gamow-Teller (GT) modes, are important probes of the nuclear forces and key ingredients for the modeling of astrophysical environments \cite{RevModPhys.75.819}. Transitions in the GT$_-$, or (p,n), branch, converting a neutron into a proton, determine $\beta^-$-decay which governs the rapid neutron-capture ("r-") process. Alternatively, transitions in the GT$_+$, or (n,p), branch can occur in core-collapse supernovae via 
electron capture on \textit{pf}- or \textit{sdg}-shell nuclei, in particular, those with a neutron number N greater than their proton number Z \cite{PhysRevC.86.015809}. 
In a simple independent-particle picture of such nuclei the transitions in the GT$_+$ channel allowed by the selection rules are, however, typically blocked by the Pauli principle due to the neutron excess. The presence of correlations in the ground state of the parent nucleus, which can smear the occupancies of the single-particle orbits, then constitutes the only mechanism that can unlock transitions in this branch.
In methods based on the mean-field approximation doubly-magic nuclei represent a test case to study these effects. Indeed, in such systems no superfluid pairing correlations are present in the mean field, and one can fully quantify
the effect of ground-state correlations (GSC) introduced beyond the mean field while avoiding possible interplays with pairing. \\
To study charge-exchange modes in large single-particle spaces one can apply the linear response theory at different levels of approximations \cite{ringschuck}. The proton-neutron Tamm-Dancoff Approximation (pn-TDA) describes such transitions as superpositions of interacting one-particle-one-hole (1p-1h) proton-neutron excitations on top of the mean-field ground state. The proton-neutron Random-Phase Approximation (pn-RPA) goes one step further and introduces additional one-hole-one-particle (1h-1p) transitions, thus generating some correlations in the parent ground state. In the following we will refer to this type of GSC as GSC$_{RPA}$. While (pn-)RPA reproduces the position of giant resonances (GRs) to a good accuracy, it is well known that it typically generates a poorly detailed description of the transition strength distributions and, in particular, is not able to describe the spreading width of the GRs.
To correct for these deficiencies one should account for higher-order configurations of the nucleons. For instance, by introducing the coupling between single nucleons and collective nuclear vibrations, the particle-vibration coupling (PVC) scheme includes, in the leading approximation, configurations of the 1p-1h $\otimes$ phonon type. \\
In the charge-exchange channel the PVC was recently implemented in consistent non-relativistic \cite{Niu2012,Niu2016,NIU2018} and relativistic \cite{Marketin2012,Litvinova2014,Robin2016,Robin2018} frameworks. These studies, however, did not include the GSC induced by the PVC effects, which we will denote as GSC$_{PVC}$ to differentiate them from GSC$_{RPA}$. In other words, configurations beyond pn-RPA were introduced in the description of states of the daughter nucleus, but not in the ground state of the parent system. It was shown in Refs. \cite{Kamerdzhiev1994,Kamerdzhiev1997}, that such GSC$_{PVC}$ can in fact be consistently included. Their effect on electromagnetic transitions was studied in a non-relativistic framework based on Landau-Migdal forces and they were found important for the description of the low-energy strength distributions. \\
In this Letter, we implement for the first time GSC$_{PVC}$ in the description of charge-exchange transitions of non-superfluid nuclei in a relativistic consistent framework. We apply the extended formalism to the calculation of GT and isovector spin-monopole (IVSM) modes in $^{90}$Zr. This nucleus is particularly interesting as it has been experimentally measured in both GT$_+$ and GT$_-$ channels up to energies beyond the GR region \cite{Yako2005,Wakasa1997}. Due to the weak pairing caused by the closed shell at $Z=50$ and sub-shell at $N=40$, $^{90}$Zr can be well approximated as doubly-magic nucleus where the effect of configuration mixing and GSC can be investigated carefully.
We will find that GSC$_{PVC}$ can induce new types of transitions from particle to particle states or from hole to hole states, that are crucial for the appearance of the GT$_+$ strength. \\
{\it Formalism.}
We want to describe the response of a nucleus to a one-body charge-changing external field, \textit{i.e.} converting a neutron (n) into a proton (p) ($\hat{F}_{-} = \sum_{pn} F_{pn} a^\dagger_p a_n$) or vice versa ($\hat{F}_{+} = ( \hat{F}_{-})^\dagger$). The corresponding transition strength distribution $S_\pm$ is fully determined by the response function $R$ describing the propagation of correlated proton-neutron pairs in the particle-hole channel:
\begin{eqnarray}
S_{\pm}(E) &=& \sum_N |\braket{\Psi_N|\hat F_{\pm}|\Psi_0}|^2 \delta (E - \Omega_N) \nonumber \\
                   &=& -\frac{1}{\pi} \lim_{\Delta\rightarrow 0^+} \mbox{Im} \langle F_{\pm}^\dagger R_{\pm}(E+i\Delta) F_{\pm} \rangle \; , 
\label{eq:strength}
\end{eqnarray}
where $\ket{\Psi_0}$ denotes the parent nuclear ground state, $\ket{\Psi_N}$ the states of the daughter nucleus, and $\Omega_N = E_N - E_0$ the corresponding transition energies.
In the following odd (resp. even) indices denote proton (resp. neutron) spherical single-particle states for the $(p,n)$ channel and \textit{vice versa} for the $(n,p)$ channel. Letters $i,j$ will be used to denote states with unspecified isospin projection. 
We introduce an index $\sigma_i$ where $\sigma_i = +$, if $i$ is a state above the Fermi level ("particle" state) and $\sigma_i = -$, if $i$ is below the Fermi level ("hole" state). The product $\sigma_{ij} \equiv \sigma_i \sigma_j$ will then be $-$ if $(i,j)$ is a particle-hole or hole-particle pair, and will be $+$ if $(i,j)$ is a particle-particle or hole-hole pair.
The response function now has, in principle, four components $R^{(--)}$, $R^{(-+)}$, $R^{(+-)}$ and $R^{(++)}$ that denote
\begin{equation}
R^{(\sigma_{12},\sigma_{34})}_{1423} \equiv R_{1423}^{\sigma_1\sigma_4\sigma_2\sigma_3} \, .
\end{equation}
That is $R^{(--)}_{1423} = \{ R_{1423}^{+--+}, R_{1423}^{-++-}, R_{1423}^{++--}, R_{1423}^{--++} \} $, $R^{(++)}_{1423} = \{ R_{1423}^{++++}, R_{1423}^{----}, R_{1423}^{+-+-}, R_{1423}^{-+-+} \} $, 
$R^{(+-)}_{1423}=  \{ R_{1423}^{+-++}, R_{1423}^{+++-}, R_{1423}^{---+}, R_{1423}^{-+--} \} $ and $R^{(-+)}_{1423} = \{ R_{1423}^{++-+}, R_{1423}^{-+++}, R_{1423}^{+---}, R_{1423}^{--+-} \} $. 
Similarly the external field has two components $F^{(-)}_{12} = \{ F^{+-}_{12}, F^{-+}_{12} \}$ and $F^{(+)}_{12} = \{ F^{++}_{12},F^{--}_{12}\}$. \\
Since in the proton-neutron (relativistic) RPA (pn-(R)RPA) nuclear excitations are superpositions of 1p-1h or 1h-1p proton-neutron transitions on top of the parent ground state, $R^{(+-)}$,  $R^{(-+)}$ and $R^{(++)}$ cancel, so that $R^{(--)}$ and $F^{(-)}$ are the only components that remain.
In order to account for higher-order configurations and describe nuclear excitations more precisely, we go beyond the pn-RRPA by introducing the coupling between single nucleons and collective vibrations of the nucleus. Up to now we have done this in the resonant Time Blocking Approximation (TBA) \cite{Tselyaev1989} by introducing PVC into the $R^{(--)}$ component via the Bethe-Salpeter equation (BSE) \cite{Robin2016,Robin2018} shown in Fig. \ref{f:Phi_NOGSC}. 
Such diagrams introduce complex configurations of the 1p-1h $\otimes$ phonon type in the states of the daughter nucleus, but do not modify the description of the parent ground state compared to pn-RRPA. The PVC, however, generates "backward-going" diagrams, as those shown in Fig. \ref{f:GSC}, which introduce complex configurations in the parent ground state. %
Such diagrams modify further the $R^{(--)}$ function of Fig. \ref{f:Phi_NOGSC} and also introduce non-zero $R^{(-+)}$, $R^{(+-)}$ and $R^{(++)}$ components of the response.
The derivation of these additional terms for non-isospin-flip transitions in a non-relativistic framework are available in \textit{e.g.} Ref. \cite{Kamerdzhiev1997}. Here we introduce them for the first time in the description of charge-exchange modes. All backward diagrams consistent with the TBA are included, and the full mathematical expressions are in the supplemental material. In the following we refer to the present extended approach as proton-neutron Relativistic Time-Blocking Approximation (pn-RTBA) with GSC$_{PVC}$. \\

\begin{figure}
\centering{\includegraphics[width=0.8\columnwidth] {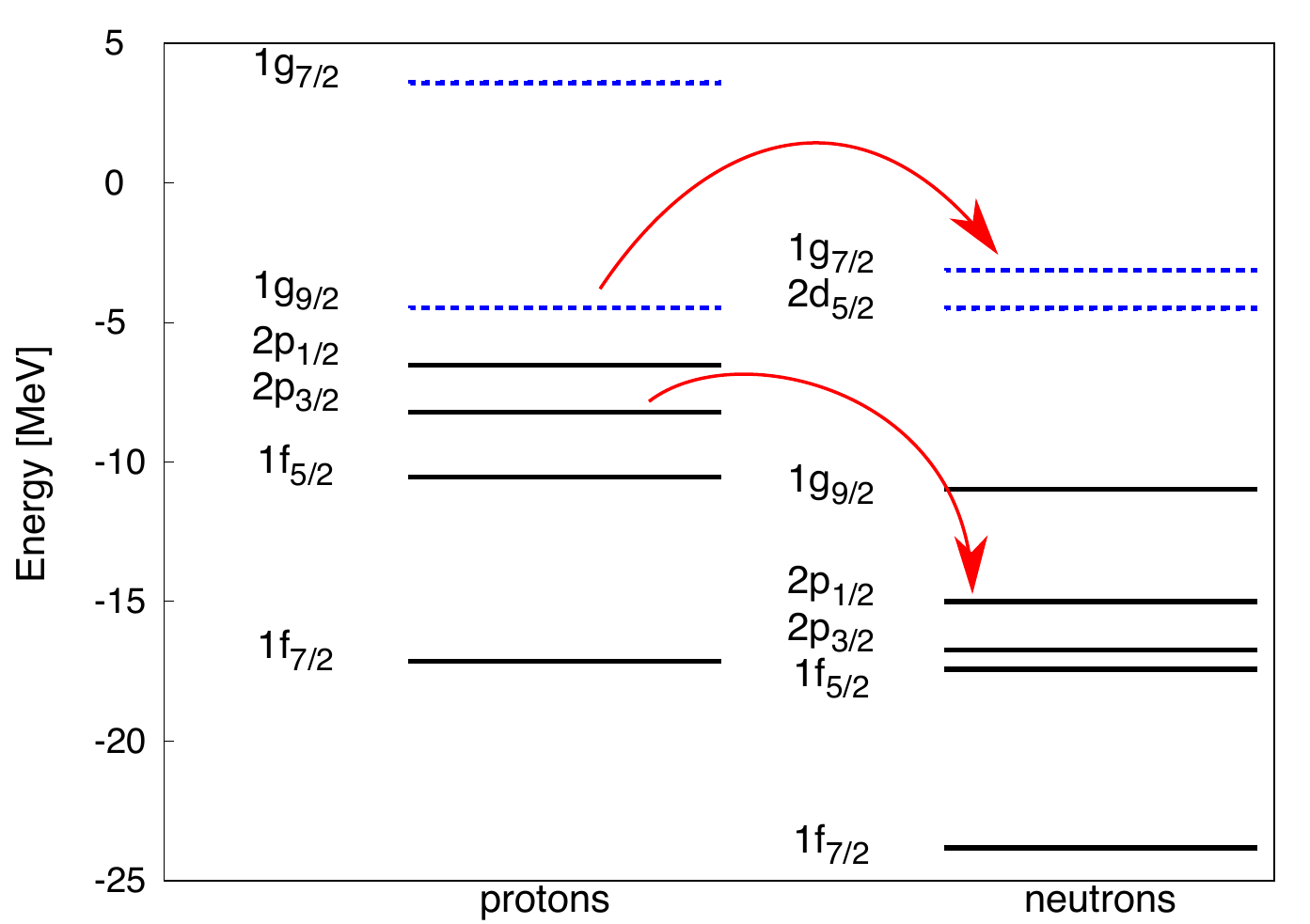}}  \hfill %
\caption{Relativistic mean-field single-particle levels in $^{90}\mbox{Zr}$ obtained with the NL3* parametrization. The full black (resp. dotted blue) levels are hole (resp. particle) states. The Fermi level is the highest hole level. The red arrows denote some transitions that are unlocked by GSC$_{PVC}$.}
 \label{f:sp_90Zr} 
\end{figure}

{\it Results.}
We apply the above formalism to the calculation of GT transitions which characterize the nuclear response to the spin-isospin-flip operator $F_{GT_{\pm}} = \sum_{i} \boldsymbol{\Sigma}^{(i)} \tau_{\pm}^{(i)}$, where $\boldsymbol{\Sigma}$ is the relativistic spin operator and $\tau_-$ (resp. $\tau_+$) converts a neutron (resp. proton) into a proton (resp. neutron). We use the NL3* parametrization of the meson-exchange interaction \cite{NL3s} and follow the numerical scheme described in Ref. \cite{Robin2018}. The pion, which does not contribute to the mean field, is included in the response with free-space coupling constant ($\frac{f_\pi^2}{4\pi}=0.08$) and associated zero-range parameter $g'=0.6$ \cite{Paar2004}.The spectrum of RRPA phonons coupled to nucleons include natural-parity neutral phonons from $0^+$ to $6^+$  and all-parity charge-exchange phonons from $0^\pm$ to $7^\pm$ with excitation energy up to 20 MeV. The single-particle states participating in the PVC are limited to a window of $50$ MeV around the Fermi levels, allowing for convergence of the strength up to this energy. \\ \\
The resulting strength distributions correspond to states in the daughter nuclei ($^{90}$Nb and $^{90}$Y). While the theoretical distributions are obtained with respect to the ground state of the parent nucleus, in order to compare to the data, it is usually necessary to relate them to the ground state of the daughter system. To this end we calculated transitions of several multipolarities ($J^\pi = 0^\pm$ to $9^\pm$) and identified the daughter ground state as the lowest peak. The energy of this state directly gives us the binding energy (BE) difference by which we should shift our GT distribution. Table \ref{t:BE} shows the corresponding BE differences calculated in the pn-RRPA, pn-RTBA, without and with the contribution of the new backward diagrams of the type shown in Fig. \ref{f:GSC}. These are compared to the experimental values \cite{Huang_2017,Wang_2017}.
\begin{table} [h!]
 \centering
\begin{tabular}{ccccc}
\hline     
                     &  EXP                                                   &pn-RRPA		&   pn-RTBA                               & pn-RTBA   \\
                     &	\cite{Huang_2017,Wang_2017}		&			& without 			                 & with  \\
                     &	                                            		&			& GSC$_{PVC} $			& GSC$_{PVC}$ \\
\hline  
\hline
 BE( $^{90}$Zr) -  BE($^{90}$Nb)  &    6.893      & 6.160 &    2.760 &   5.430                      \\   
 BE($^{90}$Zr) -  BE($^{90}$Y)  &    1.496      & 2.200     &  -1.090  &   1.880                      \\           
\hline
\end{tabular}
\caption{BE differences in MeV.}
\label{t:BE}
\end{table}
As expected pn-RRPA gives satisfactory BE differences due to the fitting of NL3* force. The inclusion of PVC without GSC worsens these values due to the lack of consistency in the description of parent and daughter nuclei.
The new GSC$_{PVC}$ induce extra binding of the parent system significantly improving the results that are brought back close to the experimental values. We also find the theoretical daughter ground states with the same angular momentum and parity as the experimental ones: $J^\pi = 8^+$ for $^{90}$Nb and $J^\pi = 2^-$ for $^{90}$Y \cite{Audi_2017}, both with and without GSC$_{PVC}$. pn-RRPA predicts $J^\pi = 5^+$ for the $^{90}$Nb ground state and $J^\pi = 2^-$ for $^{90}$Y. \\
\begin{figure}
\centering{\includegraphics[width=\columnwidth] {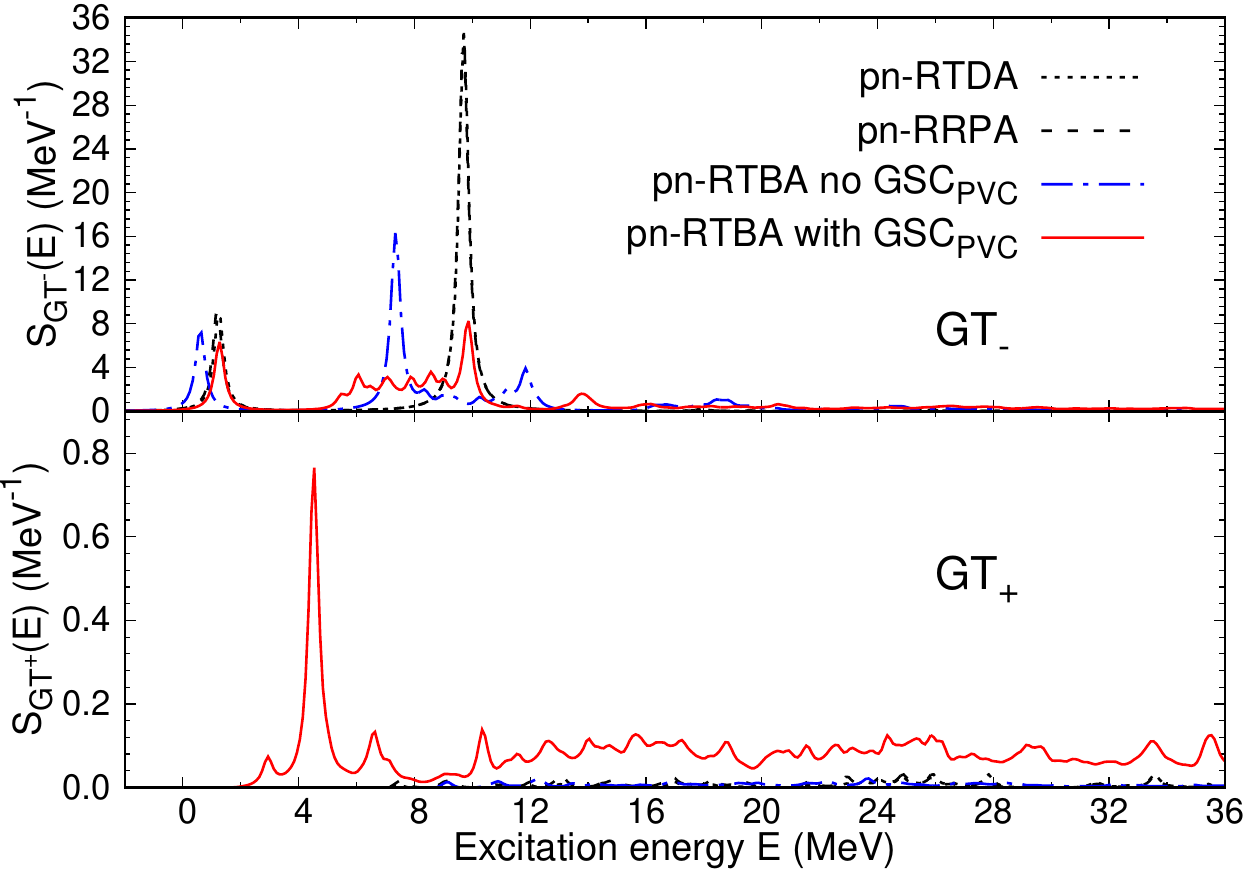}}  \hfill %
\caption{GT strength distributions $^{90}\mbox{Zr} \rightarrow ^{90}\mbox{Nb}$ (top) and $^{90}\mbox{Zr} \rightarrow ^{90}\mbox{Y}$ (bottom) calculated within pn-RTDA (dotted black), pn-RRPA (dahsed black), pn-RTBA without (dashed-dotted blue) and with (plain red) GSC$_{PVC}$. The energies are taken with respect to the theoretical daughter ground state.}
 \label{f:90Zr-200} 
\end{figure}
We show in Fig. \ref{f:90Zr-200} the GT strength distributions in $^{90}$Zr in the GT$_-$ (top panel) and GT$_+$ (bottom panel) directions. To obtain a detailed spectrum we used a smearing parameter $\Delta = 200$ keV (see Eq. (\ref{eq:strength})).
Let us first examine the GT$_-$ branch.
The dotted and dashed black curves show the results without PVC obtained at the pn-RTDA and pn-RRPA levels (i.e. without and with GSC$_{RPA}$), respectively. 
Clearly the GSC$_{RPA}$ are negligible in  $^{90}$Zr.
The blue and red curves show the results when including PVC without and with the backward-going diagrams of the type shown in Fig. \ref{f:GSC} (i.e. without and with GSC$_{PVC}$), respectively. PVC introduces fragmentation of the strength and the GSC$_{PVC}$ induce further redistribution of the GT resonance (GTR) along with an upward shift of the low-energy peak by $\sim 500$ keV.
Let us now turn to the GT$_+$ channel. In the pn-RTDA limit, the possible proton $\rightarrow$ neutron transitions respecting the GT selection rules are strongly hindered by the Pauli blocking as can be deduced from the single-particle spectrum of Fig. \ref{f:sp_90Zr}. 
Then, the GSC$_{RPA}$ can, in principle, unlock transitions from particle to hole states. However, the possible transitions appear only at excitation energies above 7 MeV and are very weak due to the small matrix elements of the external field.
The inclusion of PVC on top of pn-RRPA with only the forward-going diagrams of Fig. \ref{f:Phi_NOGSC} (blue curve) induces almost no change.
The final distribution, including the backward-going diagrams as in Fig. \ref{f:GSC}, is depicted in red.
Evidently, the GSC$_{PVC}$ have a very strong effect in the GT$_+$ channel. They induce fractional occupancies of the single-particle states of the parent nucleus, leading to new transitions from particle to particle state and from hole to hole state. In particular, the peak around $4.5$ MeV appears mainly due to transitions from the proton-$1g_{9/2}$ to the neutron-$1g_{7/2}$ and from the proton-$2p_{3/2}$ to the neutron-$2p_{1/2}$, with corresponding absolute transition densities of $0.347$ and $0.182$, respectively.
\begin{figure}[t!]
\centering{\includegraphics[width=\columnwidth] {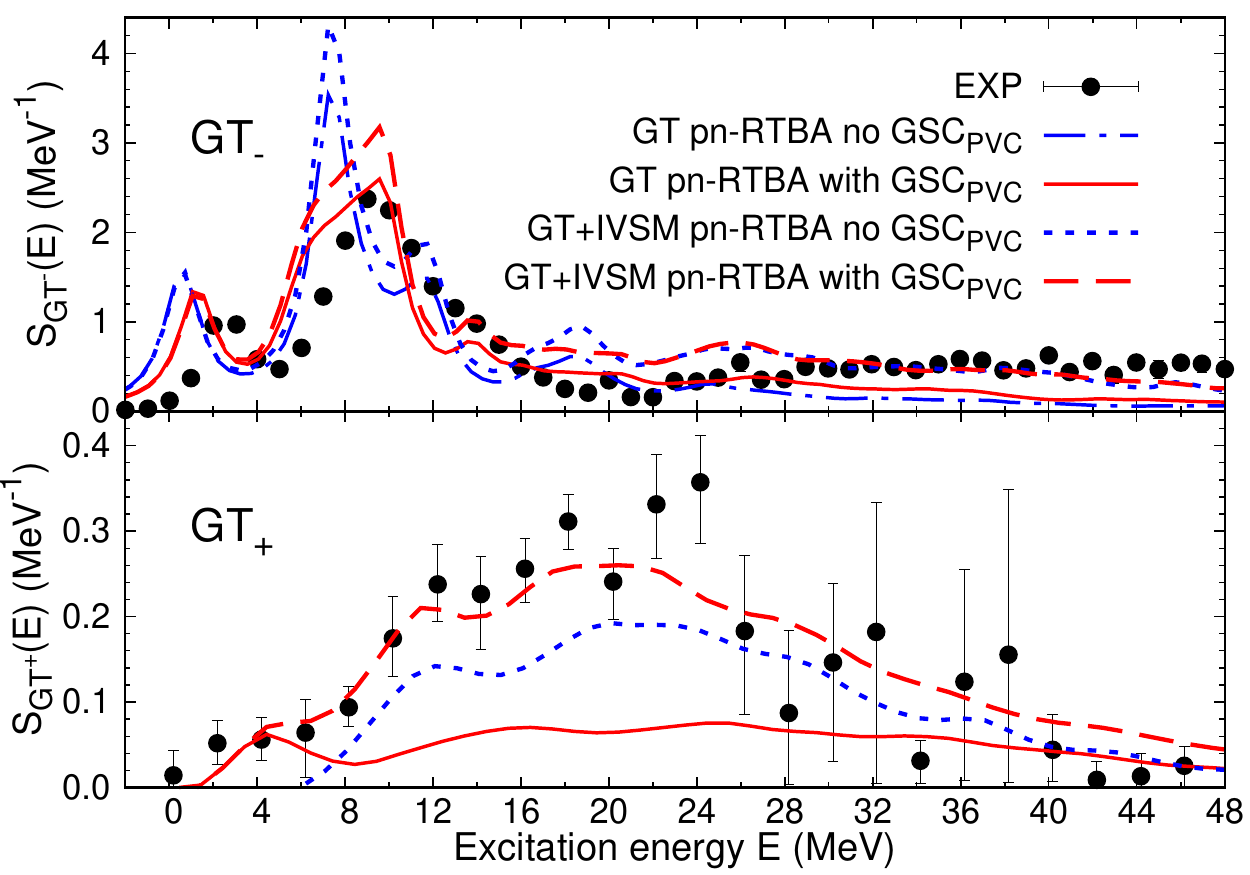}}  \hfill %
\caption{GT strength distributions $^{90}\mbox{Zr} \rightarrow ^{90}\mbox{Nb}$ (top) and $^{90}\mbox{Zr} \rightarrow ^{90}\mbox{Y}$ (bottom) smeared with $\Delta=1$ MeV and $\Delta=2$ MeV respectively. The experimental data \cite{Yako2005,Wakasa1997} is compared to the pure GT strength and the hybrid GT+IVSM strength of pn-RTBA without GSC$_{PVC}$ (dashed-dotted and dotted blue) and pn-RTBA with GSC$_{PVC}$ (plain and dashed red). }
 \label{f:90Zr} 
\end{figure}
This is illustrated in Fig. \ref{f:sp_90Zr}. \\ \\
For a comparison to the available data \cite{Yako2005,Wakasa1997}, in Fig. \ref{f:90Zr} we have smeared the calculated GT$_+$ and GT$_-$ distributions with a parameter $\Delta = 2$ and $1$ MeV, respectively, to match the experimental resolutions. 
In the GT$_-$ channel, the pn-RTBA with GSC$_{PVC}$ (plain red curve) shows a good agreement with the data up to $\sim 25$ MeV, besides a small shift of the low-energy state.
In the GT$_+$ channel, the GSC induced by PVC are solely responsible for the appearance of the low-energy peak below $6$ MeV, which is observed experimentally, as well as for higher-energy strength up to $\sim 50$ MeV. Above $\sim 6$ MeV the theoretical GT$_+$ strength alone nevertheless largely underestimates the data. It is well known, however, that at large excitation energy contributions of other multipole modes can come into play. Among them the IVSM mode, or response to the operator $F_{IVSM^{\pm}} = \sum_{i} r_{(i)}^2 \boldsymbol{\Sigma}^{(i)} \tau_{\pm}^{(i)}$, is expected to be the most important. The GT data points extracted from Refs. \cite{Yako2005,Wakasa1997} in fact also contain the contribution of such modes which could not be disentangled from the GT transitions due to the difficulty of such procedure. 
In order to have a meaningful comparison, we therefore follow the procedure of Ref. \cite{Terasaki2018}, and plot the response to the hybrid operator $F^{hyb}_{\pm} = \sum_{i} (1 + \alpha r_{(i)}^2) \boldsymbol{\Sigma}^{(i)} \tau_{\pm}^{(i)} $ where $\alpha$ is a parameter that we adjust to reproduce the magnitude of our theoretical low-energy GT strength. 
We find $\alpha = 9.1 \times 10^{-3}$ and $7.5 \times10^{-3}$ fm$^{-2}$ for the GT$_+$ and GT$_-$ branch respectively.
As seen from Fig. \ref{f:90Zr}, the IVSM mode appears responsible for the strength above 25-30 MeV in the (p,n) branch.
In the (n,p) channel it is clearly very important, even at low energy, above 5 MeV. This observation is in accordance with Ref. \cite{Miki2012}. After adding the IVSM component we obtain a very good agreement of the overall strength distribution. \\ \\
We note that, as originally discussed in Ref.~\cite{Tselyaev2007}, some of the new diagrams responsible for the GSC$_{PVC}$ (those of fourth order in the PVC vertex) can induce a small violation of the Ikeda sum rule. Numerically we find a discrepancy of $1.23 \%$ in the case of $^{90}$Zr when using a cut-off energy of 50 MeV on the nucleon states entering the PVC mechanism, and accounting for the contribution of the transitions to the Dirac sea \cite{KurasawaPRL03}. Based on a convergence study conducted in smaller single-particle bases, we expect the sum rule violation to be stable when increasing the PVC cut-off energy.
A procedure to correct for the violation by eliminating certain GSC$_{PVC}$ diagrams was proposed in Ref. \cite{Tselyaev2007}. Implementing it is beyond the scope of the present study and we leave it for a future work. \\ \\
Finally, calculations of GT modes using non-relativistic beyond-RPA methods including complex GSC were performed in the past. Ref. \cite{PhysRevC.48.1752} used a perturbative dressed pn-RPA approach and found some strength in the GT$_+$ branch of $^{90}$Zr, but only for energies above $\sim 5 $ MeV. Ref. \cite{Drozdz1990} included perturbative GSC within second RPA which produced a slight enhancement of the GTR region in $^{48}$Ca. Strong limitations of these methods on the configuration complexity, however, did not allow for a clear demonstration of the importance of complex GSC for the description of GT transitions. \\

{\it Summary and Outlook.}
We extended the pn-RTBA to consistently include the GSC arising from the coupling between single nucleons and collective nuclear vibrations. We implemented the corresponding new diagrams in a framework based on the effective meson-nucleon Lagrangian and studied their effect on the description of GT$_{-}$ and GT$_{+}$ transitions in the nucleus $^{90}$Zr which constitutes a very clear test case. The new correlations are decisive for the appearance of the GT$_+$ strength and necessary to reproduce the low-energy transitions observed experimentally in this channel. This observation is generally verified in doubly-magic nuclei with N$>$Z, where the pure independent-particle model forbids GT$_{+}$ transitions, and the GSC of pn-RPA do not allow for the appearance of such states either. Overall a very good agreement with experiment is found for $^{90}$Zr in both branches up to excitation energies of $\sim 50$ MeV when the contribution of the IVSM is accounted for. The new GSC were also necessary to reproduce the binding-energy differences with the daughter odd-odd systems.
An extension of this work to open-shell nuclei in the future will allow more systematic studies of the GSC generated by PVC and their interplay with pairing correlations. Their joint effect on beta-decay and electron-capture rates will also be investigated. 
Ultimately these developments can strongly impact astrophysical simulations which up to now use inputs from quasiparticle RPA. In the context of searches beyond the Standard Model, such as the neutrinoless double-beta decay for which an accurate description of GSC in both branches is of the utmost importance, this work could improve the predictions of RPA-based approaches and reduce the large discrepancy that exists between the many theoretical methods \cite{Engel:2016}. GSC$_{PVC}$, being one of the most difficult aspects of the quantum many-body problem, also provides a refined modeling in other areas of mesoscopic physics, such as condensed matter and quantum chemistry. 

{\it Acknowledgments.}
We thank V. I. Tselyaev and R. G. T. Zegers for enlightening discussions. This work was supported by the Institute for Nuclear Theory under US-DOE Grant DE-FG02-00ER41132, by JINA-CEE under US-NSF Grant PHY-1430152, and by US-NSF CAREER grant PHY-1654379. 
\bibliography{mybibfile}
\end{document}